\documentclass[twocolumn,showpacs,prl]{revtex4}

\usepackage{amsfonts}
\usepackage{amsthm}
\usepackage{amssymb}
\usepackage{amsmath}

\newtheorem{thm}{Theorem}

\newtheorem{lem}[thm]{Lemma}
\newtheorem{defn}[thm]{Definition}
\newcommand{\be}{\begin{equation}}
\newcommand{\ee}{\end{equation}}

\newcommand{\A}{\mathcal{A}} 
\newcommand{\B}{\mathcal{B}}

\newcommand{\X}{\mathcal{X}}

\newcommand{\rhoAB}{\rho_{\A\otimes\B}}
\newcommand{\rhoN}{\rho_{[1,n]}}
\newcommand{\rhoPN}{\rho_{1,[p,n]}}

\newcommand{\eof}[3]{E_{[#1,#2]}(#3)}
\newcommand{\bra}[1]{\langle #1|}
\newcommand{\ket}[1]{|#1\rangle}

\newcommand{\ketbra}[2]{|#1\rangle\langle #2|}
\newcommand{\pure}[1]{\ketbra{#1}{#1}}
\newcommand{\Tr}{\mathop{\mathrm{Tr}}}
\newcommand{\Ehat}{\mathbb{\hat{E}}}
\newcommand{\FCS}{\textsf{FCS}}
\newcommand{\EoF}{\textsf{EoF}}
\providecommand{\one}{\leavevmode\hbox{\small1\kern-3.8pt\normalsize1}}
\providecommand{\bysame}{\leavevmode\hbox to3em{\hrulefill}\thinspace}
\newcommand{\etal}{\textit{et al.}}

\begin{document}

\title{Entanglement in Finitely Correlated Spin States}
\author{S. Michalakis}
\email{spiros@math.ucdavis.edu}
\author{B. Nachtergaele}
\email{bxn@math.ucdavis.edu}
\affiliation{Department of Mathematics, University of California at Davis - Davis CA 95616, USA}


\begin{abstract}

We derive bounds for the entanglement of a spin with an (adjacent and
non-adjacent) interval of spins in an arbitrary pure finitely correlated state
(\FCS) on a chain of spins of any magnitude. Finitely correlated states are
otherwise known as matrix product states or generalized valence-bond states.
The bounds become exact in the limit of the entanglement of a single spin and
the half-infinite chain to the right (or the  left) of it. Our bounds provide a
proof of the recent conjecture by Benatti, Hiesmayr, and Narnhofer that their
necessary condition for non-vanishing entanglement in terms of a single spin
and the ``memory'' of the \FCS, is also sufficient \cite{BHN}. Our result also
generalizes the study of entanglement in the ground state of the AKLT model by
Fan, Korepin, and  Roychowdhury~\cite{FKR}. Our result permits one to calculate
more efficiently, numerically and in some cases even analytically, the 
entanglement of arbitrary finitely correlated quantum spin chains.
\end{abstract}

\maketitle

%%%%%%%%%%%%%%%%%%%%%%%%%%%%%%%%
%                                           				       			           %
%       				 Introduction                       			  %
%                                                                                        			  %
%%%%%%%%%%%%%%%%%%%%%%%%%%%%%%%%

Entanglement properties of quantum spin-chains have recently attracted attention from
researchers in quantum information theory and condensed matter physics. From the perspective of
quantum information theory, the distribution of entanglement over long ranges via local operations on a spin-chain~\cite{BHV, EO, NOS} has obvious applications to teleportation-based models of quantum computation
~\cite{Loc-Ent, CR, GB}. Moreover, it has recently been shown that entanglement in \textit{finitely correlated chains}~\cite{FNW2} can be used to achieve universal quantum computation~\cite{VC}
and provide a computational tool for adiabatic quantum computation~\cite{MPS}.
On the other hand, the scaling behavior of quantum correlations in infinite spin-chains is intimately related to their critical behavior (recent work has established a general mathematical framework
for studying entanglement in infinite quantum spin-chains~\cite{KMSW}.)

Motivated by the potential applications of distributed entanglement in finitely correlated chains, Benatti, \etal{} in~\cite{BHN}, 
give a {\bf necessary} condition for entanglement between a spin and a subset of other spins; namely, that the entanglement between a spin and the ``memory'' of the finitely correlated state must be non-zero.
They, furthermore, conjecture that the same condition is {\bf sufficient}, in the sense that
it implies entanglement between a spin and a subset of other spins.
We present here a proof of that conjecture by showing that the entanglement between a spin and its neighbors converges exponentially fast (in the number of neighboring spins) to the entanglement 
between a spin and the ``memory'' of the finitely correlated state. Moreover, we show that entanglement
between distant spins vanishes exponentially fast in the length of their separation.

Since finitely correlated states provide the exact ground states for generalized valence-bond solid models~\cite{Nachtergaele}, our result generalizes the calculation of entanglement~\cite{FKR} for the AKLT model~\cite{AKLT}.

More importantly, our result implies a simple and computationally efficient way for detecting distributed entanglement in finitely correlated states. Namely, the Positive Partial Transpose (PPT or Peres-Horodecki) criterion~\cite{Peres, H3} can be applied to the state describing the interactions of a spin with the "memory" of the finitely correlated state, to detect entanglement between a spin and a subset of other spins.

\section{The setup and main result} 
We will work with translation invariant pure 
{\it finitely correlated states}, \FCS~\cite{FNW2} on the infinite 
one-dimensional lattice. 
For each $i\in \mathbb{Z}$, the spin at site $i$ of the chain will be described by
the algebra $\A$ of $d\times d$ complex matrices. The observables of the spins
in an interval, $[m,n]$, are given by the tensor product $\A_{[m,n]} 
=\otimes_{j=m}^{n} (\A)_{j}$.
The algebra $\A_\mathbb{Z}$ describing the infinite chain arises as a suitable
limit of the {\it local} tensor-product algebras ${\A}_{[-n,n]}:=\otimes_{j=-n}^n(\A)_j$.
Any translation invariant state $\omega$ over $\A_\mathbb{Z}$ is completely
determined by a set of density matrices $\rho_{[1,n]}$, $n\geq 1$, which describe
the state of $n$ consecutive spins. In the case of a pure \FCS, as was
shown in \cite{FNW}, these density matrices can be constructed as follows:

The {\it memory}, $\B$, of a \FCS{} is represented by the algebra of 
$b\times b$ complex matrices. Let $\mathbb{E}:\A\otimes\B\mapsto\B$ 
be a completely positive unital map of the form
$\mathbb{E}(A\otimes B)=V(A\otimes B)V^{\dagger}$, where 
$V:\mathbb{C}^{d}\otimes\mathbb{C}^{b} \mapsto \mathbb{C}^{b}$, is a linear
map such that $VV^\dagger=\one_\B$. 
We define the completely positive map $\Ehat : \B \mapsto \B$, by $\Ehat(B) 
= \mathbb{E}(\one_{\A} \otimes B)$. The condition on $V$ implies that
$\Ehat$ is unital: $\Ehat(\one_\B)=\one_\B$.
For pure translation invariant \FCS, one can assume that there is a unique, non-singular 
$b\times b$ density matrix, $\rho$, such that $\Tr\rho\, \Ehat(B)=\Tr \rho\, B$,
for all $B\in \B$.

We introduce the density matrix $\rhoAB$ associated with the state 
encoding the interaction between the spin at site $1$ and the ``memory''
of the \FCS:
\be
{\Tr}_{\A \otimes \B}\Bigl(\rhoAB\, A\otimes B\Bigr) 
= {\Tr}_{\B}\Bigl(\rho\, \mathbb{E}(A\otimes B)\Bigr).
\nonumber
\ee
Using the cyclicity of trace we also have $\rhoAB = V^{\dagger} \rho V$.

We are now ready to define the density matrices $\rhoN$:
\be
\rhoN = {\Tr}_{\B}(V_{n}^{\dagger}\rhoAB V_{n}),\nonumber
\ee
where 
$V_{n} = (\one_{\A}\otimes V)(\one_{\A^{\otimes 2}}\otimes V) \cdots (\one_{\A^{\otimes n-1}}\otimes V)$.

An important property, intimately related to the exponential
decay of correlations in a {\it pure} \FCS{}
is that the peripheral spectrum of $\Ehat$ is trivial; 
that is, $\one_{\B}$ is the 
only eigenvector of $\Ehat$ with eigenvalue of modulus $1$ \cite{FNW}. 
This implies that the iterates of $\Ehat$ converge exponentially fast
to $\Ehat^\infty$ given by $\Ehat^{\infty}(B) = \lim_{n \rightarrow \infty} \Ehat^n(B) = \Tr(\rho B)\one_{\B}$.
More precisely, for any $\lambda$ such that $|\lambda_{i}| < \lambda < 1$,
for all eigenvalues $\lambda_{i}$ of $\Ehat$ different from $1$, there
exists a constant $c$ such that for all $n\geq 1$:
\be
\|\Ehat^{n} - \Ehat^{\infty}\| \leq c \lambda^{n}\label{Ehat:bound},
\ee
where the norm is the $\infty$-norm on $\B$ considered as a Banach space with the $1$-norm.

Our object of study is the entanglement of formation, \EoF~\cite{BDSW}.
The \EoF{} is defined for states of composite systems with 
a tensor product algebra of observables $\X_{1}\otimes\X_{2}$.
\begin{defn}[Entanglement of Formation]
The entanglement of formation of a bipartite state over $\X_{1}\otimes\X_{2}$ with associated density 
matrix $\sigma_{12}$ is given by:
\be
\eof{\X_{1}}{\X_{2}}{\sigma_{12}} = \inf \sum_{i} p_{i} \, 
S\Bigl({\Tr}_{\X_{2}}(\sigma_{12}^{i})\Bigl),
\nonumber
\ee
where $S(\rho) = - \Tr \rho \log \rho$ is the von Neumann entropy and the infimum of the average entropy is taken over all convex decompositions $\sigma_{12} = \sum_{i} p_{i}\, \sigma_{12}^{i}$ into pure states.
\end{defn}
Whenever $\X_{1}$ is finite dimensional, as will be the case for us,
the infimum can be replaced by a minimum in the above definition, i.e.,
there is an optimal decomposition, $\{p_{i},\,\sigma_{12}^{i}\}$, where the 
infimum is attained (see~\cite{EoF} for details).
We call $\{\ket{\phi_{i}}\}$ an {\bf ensemble} for the density matrix $\sigma$ whenever the latter can be
decomposed as $\sigma = \sum_{i} \pure{\phi_{i}}$.
There are an infinite number of ensembles corresponding to a given density matrix.
The following lemma provides us with a complete classification:
\begin{lem}[Isometric Freedom in Ensembles,~\cite{HJW}]\label{lem:iso-freedom}
Let $\{\ket{e_{i}}\}_{i=1}^{d}$ be the ensemble corresponding to the eigen-decomposition of
the density matrix $\sigma$, where $d = \text{rank}(\sigma)$. Then, $\{\ket{\psi_{i}}\}_{i=1}^{m}$ is
an ensemble for $\sigma$ if and only if there exists an isometry 
$U: \mathbb{C}^{d}\mapsto \mathbb{C}^{m}$
such that $$\ket{\psi_{i}} = \sum_{j=1}^{d} U_{i,j}\,\ket{e_{j}}, \, 1\leq i \leq m.$$
\end{lem}
\noindent
The above lemma implies that any two ensembles for the same density matrix,
$\{\ket{\psi_{j}}\}_{j=1}^{M_{1}}$, and $\{\ket{\phi_{i}}\}_{i=1}^{M_{2}}$,
are similarly related via a partial isometry  
$W: \mathbb{C}^{M_{1}}\mapsto\mathbb{C}^{M_{2}}$.

Our main result is the following theorem:
\begin{thm}\label{thm:conjecture}
For any pure translation invariant \FCS{} we have
\begin{align}
0 \leq \eof{\A}{\B}{\rhoAB} - \eof{\A}{\A^{\otimes n-1}}{\rhoN} \leq \epsilon(n),\label{eq:conjecture}
\end{align}
where $\epsilon(n)$ decays exponentially fast in $n$.
\end{thm}

%%%%%%%%%%%%%%%%%%%%%%%%%%%%%%%%
%                                            				       			  %
%                                    Proof of Main Theorem          			  %
%                                                                                        			  %
%%%%%%%%%%%%%%%%%%%%%%%%%%%%%%%%
\section{Proof of the Theorem}
The lower bound is proven in~\cite{BHN}. For the sake of completeness, we include here the following proof. 

The definition of $\rhoN$ implies that every decomposition of $\rhoAB$ into pure states induces
a decomposition of $\rhoN$. Moreover, the restrictions to the spin at site $1$ of the $i$-th state
in the corresponding decompositions of $\rhoAB$ and $\rhoN$ are equal. To see this, note
that since the operators $V_n$ leave the first spin invariant, the cyclicity of the trace implies
$${\Tr}_{\A^{\otimes n-1}}(\rhoN^i) = {\Tr}_{\A^{\otimes n-1} \otimes \B}(V_{n}^{\dagger} \rhoAB^i V_n) = {\Tr}_{\B}(\rhoAB^i),$$
where we have used $V_n V_{n}^{\dagger} = \one_\A \otimes \one_\B$.
It follows that for each decomposition of $\rhoAB$ there is a corresponding decomposition of $\rhoN$ with equal average entropy. Since the average entropy of $\rhoN$ is minimized over a (possibly) larger set of decompositions, the lower bound follows.

We now focus on the upper bound.
We start with the following decompositions of $\rhoAB$ and $\rhoN$ into (unnormalized) pure states:
\begin{eqnarray}
\rhoAB &=& \sum_{i=1}^{b} V^{\dagger}\ket{\chi_{i}} \bra{\chi_{i}}V\label{rhoAB:dec}\\
\rhoN &=& \sum_{i,j=1}^{b} G_{n,j}^{\dagger}V^{\dagger}\ket{\chi_{i}} \bra{\chi_{i}}V G_{n,j}, \label{rhoN:dec}
\end{eqnarray}
where $\{\ket{\chi_i}\}_{i=1}^b$ is the eigen-ensemble of $\rho$ and
$G_{n,j} = V_{n} (\one_{\A^{\otimes n}}\otimes \ket{\chi_j}/ \|\chi_j\|)$.
The term in parenthesis in the expression for $G_{n,j}$ comes from the Kraus operators in the
decomposition of the completely positive map ${\Tr_{\B}}$.

By the observation following Lemma \ref{lem:iso-freedom}, we have that 
the (unnormalized) states $\ket{\Phi^{n}_{l}}$ in the {\bf optimal decomposition}  of $\rhoN$ are given by:
\be
\ket{\Phi^{n}_{l}} = \sum_{i,j=1}^{b} U_{l,(i j)} G_{n,j}^{\dagger}V^{\dagger}\ket{\chi_{i}},\quad 1\leq l \leq L\label{optimal},
\ee
for some partial isometry $U: \mathbb{C}^{b^{2}}\mapsto \mathbb{C}^{L}$,
whose dependence on $n$ we suppress.
Moreover, it is easy to check that:
\be
\ket{\Psi_{l}} = \sum_{i,j=1}^{b} U_{l,(i j)} V^{\dagger}\ket{\chi_i},\quad 1 \leq l \leq L
\label{rhoAB:optimal},
\ee
is an ensemble for $\rhoAB$.

To calculate the \EoF{} we need the restrictions of 
$\{\pure{\Phi^{n}_{l}}\}$ and
$\{\pure{\Psi_{l}}\}$ to $\A$:
\be
\tilde{\phi}^{n}_{l} = {\Tr}_{\A^{\otimes n-1}}(\pure{\Phi^{n}_{l}}) ,\quad
\tilde{\psi}_{l} = {\Tr}_{\B}(\pure{\Psi_{l}}).\label{rhoAB:rstr}
\ee
Define the density matrices 
$\phi^{n}_{l}=\tilde{\phi}^{n}_{l}/\alpha_{l}^{n}$
and $\psi_{l}=\tilde{\psi}_{l}/\beta_{l}$, where
$
\alpha_{l}^{n} \equiv \|\tilde{\phi}_{l}^{n}\|_{1} 
= \Tr(\tilde{\phi}_{l}^{n}),\quad
\beta_{l} \equiv \|\tilde{\psi}_{l}\|_{1} \,\,= \Tr(\tilde{\psi}_{l}).
$

From the definition of the \EoF{} and the optimality of $\{\tilde{\phi}^{n}_{l}\}_{l=1}^{L}$ we get:
\be
\eof{\A}{\B \otimes \B}{\rhoAB} - \eof{\A}{\A^{\otimes n-1}}{\rhoN}
\leq \sum_{l=1}^{L}\epsilon_{l}(n)\label{eq:main},
\ee
where $\epsilon_{l}(n) = \beta_{l} S(\psi_{l}) - \alpha_{l}^{n} S(\phi^{n}_{l}).$

It remains to show that $\sum_{l=1}^{L}\epsilon_{l}(n)$ is exponentially small.
We estimate each term in the sum as:
\be
|\epsilon_{l}(n)| \leq \beta_{l} |S(\psi_{l}) - S(\phi^{n}_{l})| 
+ |\beta_{l}-\alpha_{l}^{n}| \log d\label{epsilon:first-bound},
\ee
since $\text{rank}(\phi^{n}_{l}) \leq d$.

To bound $|S(\psi_{l}) - S(\phi^{n}_{l})|$ we use Fannes' inequality for the
continuity of the von Neumann entropy~\cite{Fannes}:
\be
|S(\psi_{l}) - S(\phi^{n}_{l})| \leq (\log d+2)\|\psi_{l}-\phi^{n}_{l}\|_{1} + \eta(\|\psi_{l}-\phi^{n}_{l}\|_{1})
\label{entropy:bound},
\ee
where $\eta(x) = -x\log x$ and $\log$ is the natural logarithm.
By the triangle inequality we have:
\be
|\beta_{l}-\alpha_{l}^{n}| =
|\|\tilde{\psi}_{l}\|_{1}-\|\tilde{\phi}_{l}^{n}\|_{1}| \leq \|\tilde{\psi}_{l}-\tilde{\phi}^{n}_{l}\|_{1}\label{norms:bound}.
\ee
Another application of the triangle inequality gives:
\be
\|\psi_{l}-\phi^{n}_{l}\|_{1} 
\leq \frac{\|\beta_{l}\psi_{l}-\alpha_{l}^{n}\phi^{n}_{l}\|_{1}
+\|(\alpha_{l}^{n}-\beta_{l})\phi_{l}^{n}\|_{1}}{\beta_{l}}\nonumber,
\ee
which simplifies, with the use of~(\ref{norms:bound}), to the following inequality:
\be
\|\psi_{l}-\phi^{n}_{l}\|_{1} \leq 2\frac{\|\tilde{\psi}_{l}-\tilde{\phi}^{n}_{l}\|_{1}}{\beta_{l}}
\label{density:bound}
\ee
Combining equations~(\ref{epsilon:first-bound})-(\ref{density:bound}) and 
setting 
\be
\tau_{l}^{n} \equiv \|\tilde{\psi}_{l}-\tilde{\phi}^{n}_{l}\|_{1}/\beta_{l}, \label{tau:first-bound}
\ee
we get the following bound for $\epsilon_{l}(n)$:
\be
|\epsilon_{l}(n)| \leq \beta_{l}[(\log d^{3}+4)\tau_{l}^{n} + \eta(2\tau_{l}^{n})]
\label{epsilon:bound}.
\ee
where we have assumed that $2\tau_{l}^{n} \leq 1/e$, 
to assure $\eta(x)$ is increasing.

To complete the proof, we show that $\tau_{l}^{n}$ is exponentially
small for large $n$. Since each $G_{n,j}$ leaves the spin at site $1$ invariant, the cyclicity of the trace yields:
\begin{equation}
\tilde{\phi}^{n}_{l} = \sum_{i,i',j,j'=1}^{b}
U_{l,(i' j')}^{*}U_{l,(i j)}{\Tr}_{\B}(V^{\dagger}\ket{\chi_{i}}\bra{\chi_{i'}}V G_{n,j}G_{n,j'}^{\dagger}),
\nonumber
\end{equation}
But $G_{n,j}G_{n,j'}^{\dagger} = \one_{\A} \otimes \Ehat^{n-1}(\ket{\chi_{j}}\bra{\chi_{j'}})/(\|\chi_{j}\| \|\chi_{j'}\|)$. Substituting
$\Ehat^\infty$ for $\Ehat^{n-1}$ we get:
$$\tilde{\psi}_{l}-\tilde{\phi}^{n}_{l} = \sum_{i,i',j,j'=1}^b
U_{l,(i' j')}^{*} U_{l,(i j)} {\Tr}_{\B}(X_{i,i'}Y_{j,j'}),$$
where $X_{i,i'} = V^{\dagger}\ket{\chi_{i}}\bra{\chi_{i'}}V$ and
$Y_{j,j'} = \one_{\A} \otimes [\Ehat^{n-1}-\Ehat^{\infty}](\ket{\chi_{j}}\bra{\chi_{j'}})/(\|\chi_{j}\| \|\chi_{j'}\|)$.

Like all trace preserving quantum operations, the partial trace is contractive with respect to the $1$-norm. Hence, an application of the triangle inequality for the $1$-norm gives:
\begin{eqnarray*}
\|\tilde{\psi}_{l}-\tilde{\phi}^{n}_{l}\|_{1} \leq \sum_{i,i',j,j'}^b
|U_{l,(i' j')}^{*}| |U_{l,(i j)}| \|X_{i,i'}\|_1 \|Y_{j,j'}\|_1 \label{density:2-norm}
\end{eqnarray*}
It is not hard to see that
\be
\|X_{i,i'}\|_1 = \|\chi_{i}\| \|\chi_{i'}\|, \quad  \|Y_{j,j'}\|_1 \leq \|\Ehat^{(n-1)}-\Ehat^{\infty}\| \nonumber
\ee
and hence
\be
\|\tilde{\psi}_{l}-\tilde{\phi}^{n}_{l}\|_{1} \leq \Bigl(\sum_{i=1}^{b}
\Bigl |\sum_{j=1}^b U_{l,(i j)}\Bigr | \, \|\chi_{i}\|\Bigl)^{2}\,\|\Ehat^{(n-1)}-\Ehat^{\infty}\| \nonumber
\ee
Since $\sum_{i,j=1}^{b} |U_{l,(i j)}|^{2}\, \|\chi_{i}\|^{2} = \beta_{l}$, two applications of Cauchy-Schwarz give:
\be
\|\tilde{\psi}_{l}-\tilde{\phi}^{n}_{l}\|_{1} \leq b^2 \beta_{l} \|\Ehat^{(n-1)}-\Ehat^{\infty}\|.
\label{density:2-norm-final}
\ee

Finally, combining~(\ref{Ehat:bound}) with~(\ref{density:2-norm-final}), equation~(\ref{tau:first-bound}) becomes:
\be
\tau_{l}^{n} \leq c_{1} \lambda^{n}, \quad c_{1} = c b^2 / \lambda.\label{tau:final-bound}
\ee

To conclude the proof, we note that since the bound for $\tau_{l}^{n}$ is independent of $l$, summing over $l$ in equation~(\ref{epsilon:bound}) yields:
\be
\sum_{l=1}^{L} |\epsilon_{l}^{n}| \leq (\log d^{3}+4) c_{1}\lambda^{n} + \eta(2 c_{1}\lambda^{n}).
\nonumber
\ee

It is clear that for $\lambda' > \lambda$ there exists a constant $c_{2}$ such that 
$$\eta(2 c_{1}\lambda^{n}) \leq c_{2}(\lambda')^{n}.$$
The only condition on $n$ was imposed in equation~(\ref{epsilon:bound})
were we assumed that $2\tau_{l}^{n} \leq \frac{1}{e}$. Using equation~(\ref{tau:final-bound}) we see 
that there is an $n_{0}$ such that the above condition is satisfied for all $n \geq n_{0}$.
The previous observations imply that for all $\lambda'$ with $\lambda < \lambda' < 1$, 
there is a constant $c_{3}$ such that:
\be
\epsilon(n) = c_{3} (\lambda ')^{n} \geq \sum_{l=1}^{L} |\epsilon_{l}^{n}|,\text{  for all $n$.}\nonumber
\ee
Finally, equation~(\ref{eq:main}) implies that:
\be
\eof{\A}{\B \otimes \B}{\rhoAB} - \eof{\A}{\A^{\otimes n-1}}{\rhoN} \leq \epsilon(n)\nonumber,
\ee
and this completes the proof of the theorem.

A natural question to ask at this point is the following: 
\emph{How does the entanglement between the spin at site $1$
and spins at sites $[p,n],\, p\geq 2$ behave as $p$ becomes large?}
Since the state $\rhoPN$ factorizes into $\rho_{1}\otimes \rho_{[p,n]}$
as $p \rightarrow \infty$~\cite{HP}, we expect that the bulk of the entanglement  is
concentrated near site $1$.
The following theorem confirms this:
\begin{thm}\label{thm:distant-ent}
For any pure translation invariant \FCS{} and $n \geq p \geq 2$, the following bound holds:
\be
\eof{\A}{\A^{\otimes n-p+1}}{\rhoPN} \leq \epsilon(p),\label{eq:distant-ent}
\ee
where $\epsilon(p)$ decays exponentially fast in $p$.
\end{thm}
\begin{proof}[Sketch of the proof:]
The main observation is that the trace distance between the states $\rhoPN$ and $\rho_1\otimes\rho_{[p,n]}$ vanishes exponentially fast with $p$. This is a consequence of the exponential rate of
convergence described in equation~(\ref{Ehat:bound}). Since
$\eof{\A}{\A^{\otimes n-p+1}}{\rho_1\otimes\rho_{[p,n]}} = 0$,
a straightforward application of Nielsen's inequality for the continuity of the \EoF~\cite{Nielsen}
yields the desired result.
\end{proof}

\section{Discussion}
Having established such a strong connection between the states $\rhoN$ and $\rhoAB$, one can
apply various entanglement criteria on $\rhoAB$ to deduce entanglement properties of the spin chain.
To start with, we note that for qubit chains with $2$-dimensional memory algebra $\B$, the entanglement
of $\rhoN$ can be computed analytically (in the limit) by evaluating the concurrence~\cite{Wooters} of
$\rhoAB$. For higher dimensions one can apply the PPT criterion to 
$\rhoAB$ to detect distributed entanglement in the finitely correlated state. Specifically, the main theorem in~\cite{Horodecki} implies that there can be no PPT bound entanglement in $\rhoAB$ since
rank$(\rhoAB)=b \leq \max\{d,b\}$. Hence, if the partial transpose of $\rhoAB$ is positive, then $\rhoAB$
is separable. On the other hand, if the partial transpose of $\rhoAB$ is negative, then for $n$ large
enough $\rhoN$ becomes entangled. The amount of maximum entanglement in $\rhoN$ depends on
the amount of entanglement found in $\rhoAB$. From this point of view, it would be very interesting to
look at \FCS{} that maximize entanglement of $\rhoAB$. Moreover, understanding how entanglement of $\rhoAB$ varies with different CP maps $\mathbb{E}$ could lead to a better understanding of how phase
transitions occur when we vary the parameters in the underlying hamiltonian of the system.

To conclude, we note that the conjecture of Benatti, \etal~\cite{BHN}, 
follows as a corollary of Theorem~\ref{thm:conjecture}. In particular, our result implies that a spin at site $1$ of the chain is entangled
with spins at sites $[2,n]$ (for $n$ large enough) if and only if $\rhoAB$ is entangled. Moreover,
the entanglement of $\rhoN$ approaches the entanglement of $\rhoAB$ exponentially fast.
\begin{acknowledgments}
S. Michalakis would like to thank F. Benatti, H. Narnhofer and P. Horodecki for valuable discussions.
The authors acknowledge support from NSF Grants DMS-0303316 and DMS-0605342.
\end{acknowledgments}


\begin{thebibliography}{0}
\bibitem{BHN}
F. Benatti, B.C. Hiesmayr, and H. Narnhofer,
Europhys. Lett. {\bf 72} (1), 28-34 (2005).

\bibitem{FKR}
H. Fan, V. Korepin, and V. Roychowdhury,
Phys. Rev. Lett. {\bf 93}, 227203 (2004) 

\bibitem{BHV}
Bravyi, S. and Hastings, M.B. and Verstraete, F.,
arXiv:quant-ph/0603121.

\bibitem{EO}
Eisert, J. and Osborne, T.J.,
arXiv:quant-phys/0603114.

\bibitem{NOS}
Nachtergaele, B. and Ogata, Y. and Sims, R.,
J. Stat. Phys., to appear, arXiv:math-ph/0603064.

\bibitem{Loc-Ent}
M. Popp, F. Verstraete, M. A. Mart'n-Delgado, and J. I. Cirac,
Phys. Rev. A {\bf 71}, 042306 (2005)

\bibitem{CR}
L. Campos Venuti, and M. Roncaglia,
Phys. Rev. Lett. {\bf 94}, 207207 (2005)

\bibitem{GB}
V. Giovannetti, and D. Burgarth,
Phys. Rev. Lett. {\bf 96}, 030501 (2006) 

\bibitem{FNW2}
M. Fannes, B. Nachtergaele, and R.F. Werner, 
Comm. Math. Phys. {\bf 144}, 443-490 (1992)

\bibitem{VC}
F. Verstraete, and I. Cirac,
Phys. Rev. A {\bf 70}, 060302(R) (2004).

\bibitem{MPS}
M. C. Banuls, R. Orus, J. I. Latorre, A. Perez, P. Ruiz-Femenia,
Phys. Rev. A {\bf 73}, 022344 (2006)

\bibitem{KMSW}
M. Keyl, T. Matsui, D. Schlingemann, and R.F. Werner,
math-ph/0604071 (2006)

\bibitem{Nachtergaele}
B. Nachtergaele,
Commun. Math. Phys., {\bf 175}, 565 (1996)

\bibitem{AKLT}
I. Affleck, T. Kennedy, E. H. Lieb, H. Tasaki,
Phys. Rev. Lett. {\bf 59} 799 (1987);
\bysame,
Commun. Math. Phys., {\bf 115} 477 (1988)

\bibitem{Peres}
A. Peres,
Phys. Rev. Lett. {\bf 77}, 1413 (1996)

\bibitem{H3}
M. Horodecki, P. Horodecki, R. Horodecki,
Phys. Lett. A {\bf 210} (1996)

\bibitem{FNW}
M. Fannes, B. Nachtergaele, and R.F. Werner, 
Journal of Functional Analysis {\bf 120}, 511-534 (1994)

\bibitem{BDSW}
C.H. Bennett, D.P. DiVincenzo, J.A. Smolin, and W.K. Wooters,
Phys. Rev. A {\bf 54}, 3824 (1996).

\bibitem{EoF}
A. W. Majewski,
J. Phys. A: Math. Gen. {\bf 35}, 123-134 (2002)

\bibitem{HJW}
L. P. Hughston, R. Jozsa and W. K. Wootters,
Phys. Lett. A {\bf 183} (1), pp. 14-18 (1993).

\bibitem{Fannes}
M. Fannes, 
Comm. Math. Phys. {\bf 31}, 291-294 (1973)

\bibitem{HP}
F. Hiai, D. Petz, 
Journal of Functional Analysis {\bf 125}, 287-308 (1994)

\bibitem{Nielsen}
M. A. Nielsen,
Phys. Rev. A {\bf 61}, 064301 (2000).

\bibitem{Wooters}
S. Hill, W. K. Wooters,
Phys. Rev. Lett. {\bf 78}, 5022 (1997).

\bibitem{Horodecki}
P. Horodecki, M. Lewenstein, G. Vidal, I. Cirac,
Phys. Rev. A {\bf 62}, 032310 (2000).

\end{thebibliography}
\end{document}